\documentclass{ws-ijmpe_2}
\usepackage{graphicx,epstopdf}
\usepackage[english]{babel}

\title{
\includegraphics[width=0.35\textwidth]{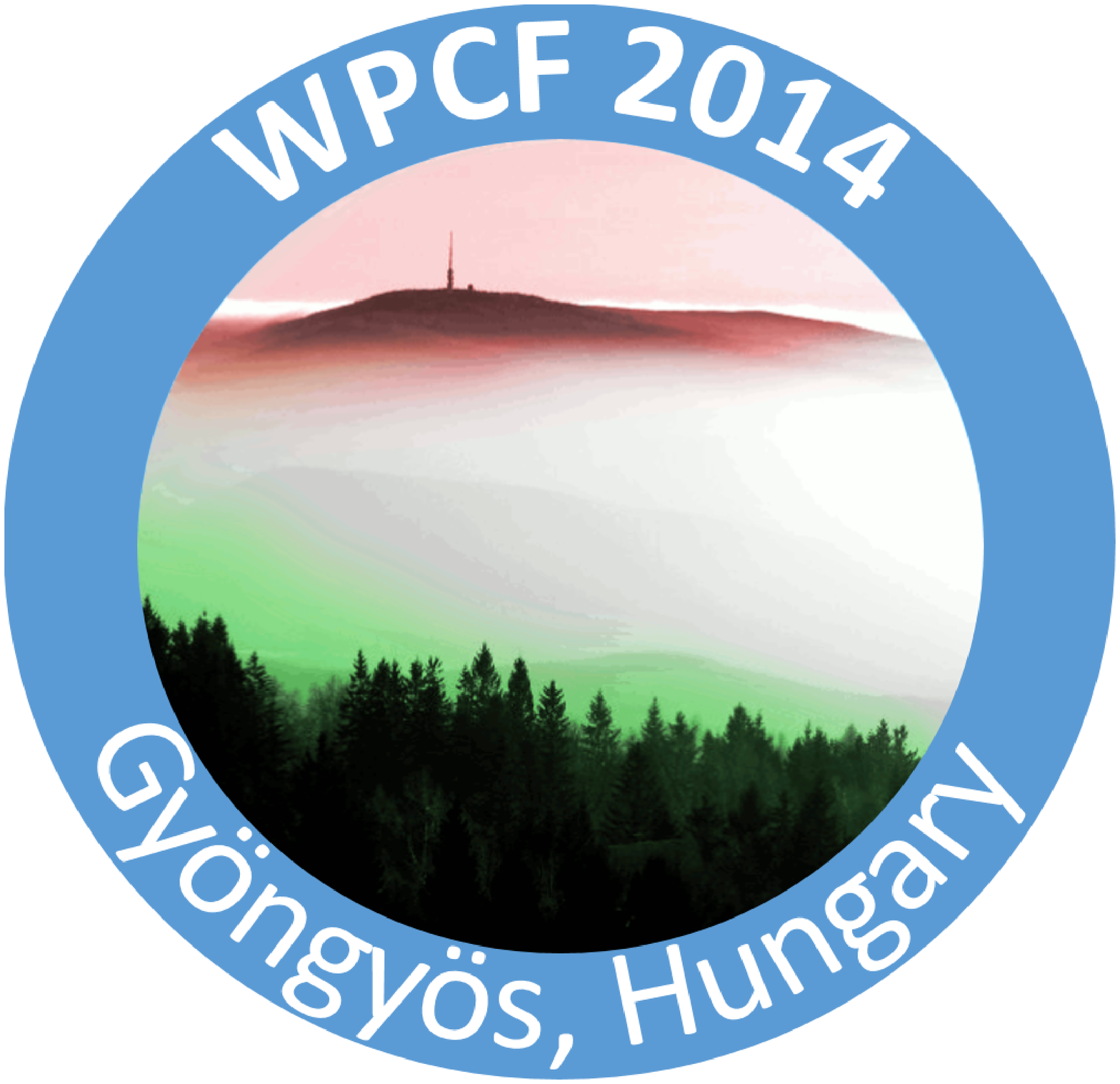}\\[1cm]
Three-component Pomeron model in high energy  $pp-$ and $\overline{p}p-$ elastic scattering}
\author{{A. Lengyel, Z. Tarics}\\[1ex]
Institute of Electron Physics, Ukrainian Nat. Ac. of Sci. \\
Universitetska, 21, Uzhgorod, Ukraine\\
}

\begin{document}

\fontfamily{lmss}\selectfont
\maketitle

\begin{abstract}
We assume that the Pomeron is a sum of Regge multipoles, each
corresponding to a finite gluon ladder. From the fit to the data
of $pp-$ and $\overline{p}p-$ scattering at high energy and all
available momentum transfer we found that taking into
account the spin, three-term multipole Pomeron and Odderon with different form-factors
 are substantial for good description of
differential, total cross section and ratio $\rho$ in the whole high
energy experimental domain.
\end{abstract}

\section{Introduction}
%\label{s0}
The Pomeron being an infinite gluon ladder \cite{FKL}-\cite{FL98} may appear
as a finite sum of gluon ladders corresponding to a finite sum of
Regge multipoles with increasing multiplicities \cite{Fiore}-\cite{UFZ}. The
first term in $\ln s$ series contributes to the total
cross-section with a constant term and can be associated with a
simple pole, the second one (double pole) goes as $\ln s$, the
third one (triple pole) as $ln^{2} s$, etc. All Pomeron poles have
unit intercept. Previously the multipole Pomeron and many-Pomerons approaches were
 investigated on different applications \cite{Giffon}-\cite{Kontros} (see also review \cite{Dremin}).
 Due to the recent experiments on elastic and inelastic proton-proton scattering by the TOTEM Collaboration at the LHC \cite{7tev2}, data in a wide range, from lowest up to TeV energies,  both for proton-proton and antiproton-proton scattering in a wide span of transferred momenta are now available.
The experiments at TeV energies give a chance to verify different Pomeron and Odderon models because the secondary Reggeon contributions at these energies are small.
Note that none of the existing models of elastic scattering was able to predict the value of the differential cross section beyond the first cone, as clearly seen in Fig.4 of the TOTEM paper \cite{7tev2,7tev1}.  This problem is still remains actual and
inspired some people to construct the new and revitalize a well-known old fashioned models.
Consequently for this time it is necessary to revise our concept on exchange mechanism in this  distant region of energies
\cite{landshoff}-\cite{Godizov}. A number of models have been refined and developed \cite{JLL}-\cite{Pancheri}.
In fact, several of used Pomeron models, as a rule, have intercept $>1$ which requires the unitarisation \cite{prokudin}, other ones have complicated structure and the overall number of parameters was fairly high \cite{Menon,martynov}.

Here we suggest the three-component Pomeron
model inspired by the finite sum of gluon ladders extended to the
whole range of available momentum transfer of high-energy $pp-$
and $\overline{p}p-$ elastic scattering and performed a
simultaneous fit to the $\sigma _{tot}$, $\rho$ and $d\sigma /dt$
data.
Our goal is to investigate the capabilities of Pomeron model as a finite sum of
gluon rungs ($\ln s$ power) equivalent to single pole + dipole +
tripole Pomeron with sufficiently account the spin influence and non-linear trajectory for $pp-$
and $\overline{p}p-$ scattering in first and second diffraction cone.
Our strategy is two fold: one to select a core set of experimental data, as well as models of reference,
most appropriately describing the details of this basic set. First we take the set \cite{data}, and then - the paper \cite {martynov}.

This paper is organized as follows. In the next section, one
introduces the main formulas and features of the model. In Sec. 3,
we perform the comparison with experiment. In the last section, the
conclusions are drawn up.
\section{The model}

The reduced form of nucleon-nucleon amplitude  without double spin-flip accounting is \cite{ter}:
\begin{equation}
%\label{eq1}
A(s,t )=A_{00}(s,t )+ \frac{\sqrt{-t}}{2m_p}  \frac{}{}   A_{01}(s,t ),
\end{equation}
where $A_{00}(s,t )$ is spin-nonflip component and $A_{01}(s,t )$ is spin-flip component of scattering amplitude.

Our ansatz for the spin-nonflip scattering amplitude component $A_{00}(s,t)$  is:
\begin{equation}
%\label{eq1}
A_{00}(s,t)=P_{00}(s,t)+R_f(s,t)\pm
\left[R_{\omega}\left(s,t\right)+ O\left(s,t\right)\right]
\end{equation}
for $\overline{p}p$ (upper symbol) and ${pp}$ (lower symbol) scattering respectively.
For the spin-flip scattering amplitude component $A_{01}(s,t)$
\begin{equation}
%\label{eq2}
A_{01}=P_{01}(s,t)+R_f(s,t)\pm \left[R_{\omega}\left(s,t\right)+g_{od} O\left(s,t\right)\right].
\end{equation}
We suppose that the contribution of the subleading reggeons to the spin-flip amplitude is the same as to the spin-nonflip amplitude and the contribution of Odderon to the spin-flip amplitude differ from one to the spin-nonflip amplitude by the factor  $g_{od}$, where the Pomeron contribution we introduce in form:
\begin{equation}
%\label{eq3}
P_{00}\left(s,t\right)= is {\left(-i\frac{s}{s_0}\right)}
^{\alpha_{P}(t)-1}\sum_{j=0}^2 a_{0j}\ln^j
\left(-i\frac{s}{s_0}\right)e^{\varphi_{0j}(t)}
\end{equation}
and
\begin{equation}
%\label{eq4}
P_{01}\left(s,t\right)= is {\left(-i\frac{s}{s_0}\right)}
^{\alpha_{P}(t)-1}\sum_{j=0}^2 g_{1j}\ln^j
\left(-i\frac{s}{s_0}\right)e^{\varphi_{1j}(t)}.
\end{equation}

The Pomeron trajectory is:
\begin{equation}
%\label{eq5}
\alpha_{P}(t)=1+ \alpha_P^{\prime}t+\alpha_P^{\prime \prime}\left(\sqrt{t_{\pi}}-\sqrt{{t_{\pi}}-{t}}\right),
\end{equation}

where the lowest two-pion threshold $t_{\pi}=4m_{\pi}^{2}$. The
residue functions are:

\begin{equation}
%\label{eq6}
\varphi_{0j}(t)=\gamma_{0j}\left(\sqrt{t_{\pi}}-\sqrt{{t_{\pi}}-{t}}\right),
\end{equation}
\begin{equation}
 \varphi_{1j}(t)=\gamma_{1j}\left(\sqrt{t_{\pi}}-\sqrt{{t_{\pi}}-{t}}\right).
\end{equation}

In %(\ref{eq7})
(2),(3) the $R_f(s,t),R_\omega(s,t)$ and $O(s,t)$ contain the
subleading reggeons as well as the Odderon contributions to the
scattering amplitude:
\begin{equation}
%\label{eq8}
R_f\left( s,t\right) =g_f\left( -i\frac s{s_0}\right)
^{\alpha _f\left( t\right) }e^{b_ft}
\end{equation}
and
\begin{equation}
% \label{eq9}
 R_{\omega}\left( s,t\right) =ig_{\omega}\left( -i\frac
s{s_0}\right) ^{\alpha _{\omega}\left( t\right) }e^{b_{\omega}t},
\end{equation}

 where
\begin{equation}
%\label{eq7}
\alpha _j\left( t\right) =1+\alpha _j^{^{\prime }}t,
\quad j=f,\omega; \quad s_{0}=1~ GeV^{2}.
\end{equation}
To describe the different behavior of proton-proton and
antiproton-proton differential cross-section in region of dip-bump one
needs to include the Odderon contribution, which we use in a
simple form:

\begin{equation}
%\label{eq8}
O\left(s,t\right)= s{\left(-i\frac{s}{s_0}\right)}
^{\alpha_{O}(t)-1}\sum_{j=0}^2 g_j\ln^j
\left(-i\frac{s}{s_0}\right)e^{\phi_{j}(t)}.
\end{equation}

The Odderon trajectory is

\begin{equation}
%\label{eq9}
\alpha_{O}(t)=1+  \alpha _O^{{\prime }}t.
%+\alpha _O^{\prime \prime}\left(\sqrt{t_{\pi}}-\sqrt{{t_{\pi}}-{t}}\right),
\end{equation}
 The residue functions are:
\begin{equation}
\label{eq10}
\phi_{j}(t)=\delta_{j}\left(\sqrt{t'_{\pi}}-\sqrt{{t'_{\pi}}-{t}}\right),
\end{equation}

$t'_{\pi}=9m_{\pi}^{2}$.

\begin{table}
\caption {Parameters, quality of the fit  obtained in the whole interval in $s$ and $t$. }
\vskip3mm
\label{tab:fitparam}
\tabcolsep18.2pt
\begin{center}\small
\begin{tabular}{|lcc|} \hline
Parameter & Value & Error\\
\hline
\hline
$a_{00}$ & 8.449 & 0.121 \\
    \hline
  $a_{01}$ & -0.855 & 0.0188 \\
  \hline
  $a_{02}$ & 0.06519 & $0.81.10^{-2}$ \\
        \hline
  $a_{10}$ &-0.3690 & 0.0250 \\
    \hline
  $a_{11}$ & 6.134 & 0.0910 \\
  \hline
  $a_{12}$ & -0.1938 & 0.00461 \\
  \hline
  %$\alpha^{\prime}_P,GeV^{-2}$ &  0.4 & fixed \\
  %\hline
  $\alpha^{\prime\prime}_P,GeV^{-1}$ & 0.04638 & $0.523.10^{-2}$ \\
  \hline
  $\gamma_{00},GeV^{-1} $ & 2.107 & 0.037 \\
  \hline
  $\gamma_{01},GeV^{-1}$ & 0.5742 & 0.0562 \\
  \hline
  $\gamma_{02},GeV^{-1}$ & 1.422 & 0.0633 \\
  \hline
  $\gamma_{10},GeV^{-1} $ & 2.623 & 0.0515 \\
  \hline
  $\gamma_{11},GeV^{-1}$ & 6.540 & 0.02655 \\
  \hline
  $\gamma_{12},GeV^{-1}$ & 5.176 & 0.0354 \\
  \hline
  $ g_{00}$ & -0.4855 & 0.0188 \\
    \hline
  $ g_{01}$ & -0.4855 & 0.0188 \\
    \hline
  $ g_{02}$ & 0.06519 & $0.81.10^{-3}$ \\
    \hline
 $\alpha^{\prime}_O,GeV^{-2}$ &  0.05197 & $0.117.10^{-2}$ \\
  \hline
 $ \delta_{0},GeV^{-1}$ & 2.623 & 0.0515 \\
    \hline
  $ \delta_{1},GeV^{-1}$ & 6.450 & 0.026 \\
    \hline
  $ \delta_{2},GeV^{-1}$ & 5.176 & 0.035 \\
    \hline
  $ g_{od}$ &553.6 & 47.7 \\
    \hline

  $a_{f}$ & -10.75 & 0.75 \\
  \hline
  $\alpha_{f}$ &  0.5395 & 0.0185 \\
  \hline
  % $ b_f GeV^{-2}$ & 12.0 & fixed  \\
  % \hline
  $ a_{\omega}$ & 10.17 & 0.44 \\
  \hline
  $\alpha_{\omega}$ &  0.4182 & 0.0131 \\
  \hline
 %$ b_{\omega}, GeV^{-2}$ & 14 & fixed  \\
 %\hline
$\chi^2/dof$ & 1.56 & \\
\hline
%$\sigma_{tot}(7 TeV) $& $98.1\pm 0.1 $ & \\
%\hline
%$\sigma_{tot} (14 TeV)$ & $111.4\pm 0.1$ & \\
%\hline
\end{tabular}
\end{center}
\end{table}
\section{Comparison with experiment}

In order to determine the parameters that control the
$s$-dependence of $A\left(s,0\right)$ we applied a wide energy
range $5GeV\leq \sqrt{s}\leq 7000 GeV$ and used the available data for total
cross sections and $\rho$ ( \cite{data}). A
total of $107$ experimental points were included for $t=0$. For
the differential cross sections we selected the data at the
energies $\sqrt{s}=19; 23; 31; 44; 53; 62; 7000 GeV$ (for $pp-$scattering)(1633 experimental points)
and $\sqrt{s}$=31;53;62;546;1800$ GeV$ (510 experimental points) for
 $\overline{p}p-$ scattering \cite{data}. The squared 4-momentum covers entire available range
$0.01(GeV)^2{<|t|<}14(GeV)^2$. The grand total number of
2384 experimental points were used in overall fit.
In the calculations we use the following normalization for the dimensionless amplitude:
\begin{equation}
%\label{eq11}
\sigma_{tot}=\frac{4\pi}s ImA\left(s,t=0\right),
\end{equation}

\begin{equation}
\quad\frac{d\sigma}{dt}=
\frac{\pi}{s^2}[
\left|A_{00}(s,t)\right|^2-\frac{t}{{4m_p}^2}\left|A_{01}(s,t)\right|^2].
\end{equation}

The resulting fits for $\sigma _{tot}$, $\rho$, $\frac{d\sigma }{dt}$
are shown in  Figs. 1. - 2. with the values of the fitted parameters
quoted in Table 1. From these figures we conclude that the
multipole Pomeron model corresponding to a sum of gluon ladders up
to two rungs complicated with Odderon contribution and spin counting fits the data well in a wide energy and momentum transfer regions.
In this paper, we have explored only the simplest phenomenological tripole Pomeron. In fact, the scattering amplitude is much more complicated than just a simple power series in $\ln s$. On the one hand, although we used just a simplified $t$-dependence in the model, reasonably good results were obtained. Because the slopes of secondary reggeons do not influence the fit sufficiently, we
have fixed them at $\alpha_{f}'=0.84 (GeV)^{-2}$ and $\alpha_{\omega}'=0.93 (GeV)^{-2}$, which correspond to the values of Chew-Frautschi plot, as well as its slope parameters $b_{f}=12.0 (GeV)^{-2}$ and $b_{\omega}=14.0 (GeV)^{-2}$. Additionally we fixed the Pomeron trajectory slope $\alpha_{P}^\prime =0.4(GeV)^{-2}$. On the other hand, we included the
curvature of the Pomeron trajectory that cannot be negligible. The quality of our fits $\chi^2/dof$=1.56 is comparable with that of the best fit of \cite{martynov}.

\section{Summary}

We have approved the tripole Pomeron model
 having each term corresponding to a finite
gluon ladder.
% instead of a factor corresponding to the
%supercritical Pomeron and conserving other useful properties of
%the model.

This corresponds to the finite sum of gluon ladders
with up to two rungs or alternatively up to the tripole Pomeron
contribution. We have obtained very good description of $pp$ and
$\overline{p}p$ hadron scattering data at intermediate and high
energies and all available momentum transfer. We conclude that the nonfactorisable
form of the Pomeron and Odderon amplitudes as well as the
nonlinearity  of its trajectory and the residue function is
strongly suggested by data at all available momentum transfer.
It should be noted that the addition of spin-flip component of scattering amlitude decisively improves the result of the fit.

\section*{Acknowledgments}
We are  grateful to L.Jenkovszky and E.Martynov for valuable suggestions and A.M.Lapidus for fruitful discussions.

\begin{figure}[fig:Pomeron2]
\center{
        \includegraphics[width=0.45\linewidth]{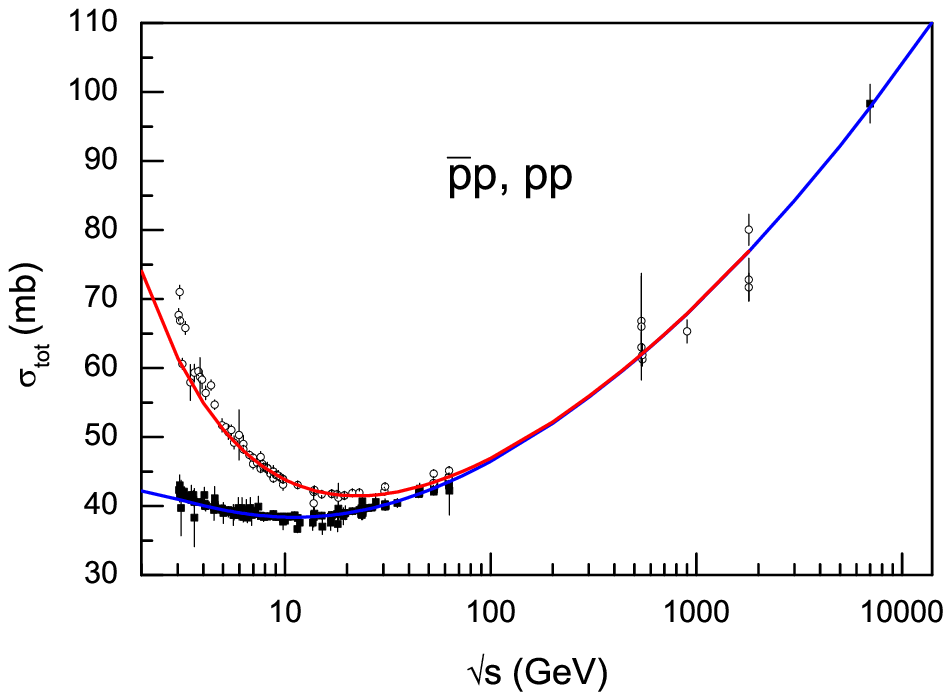}
        \includegraphics[width=0.45\linewidth]{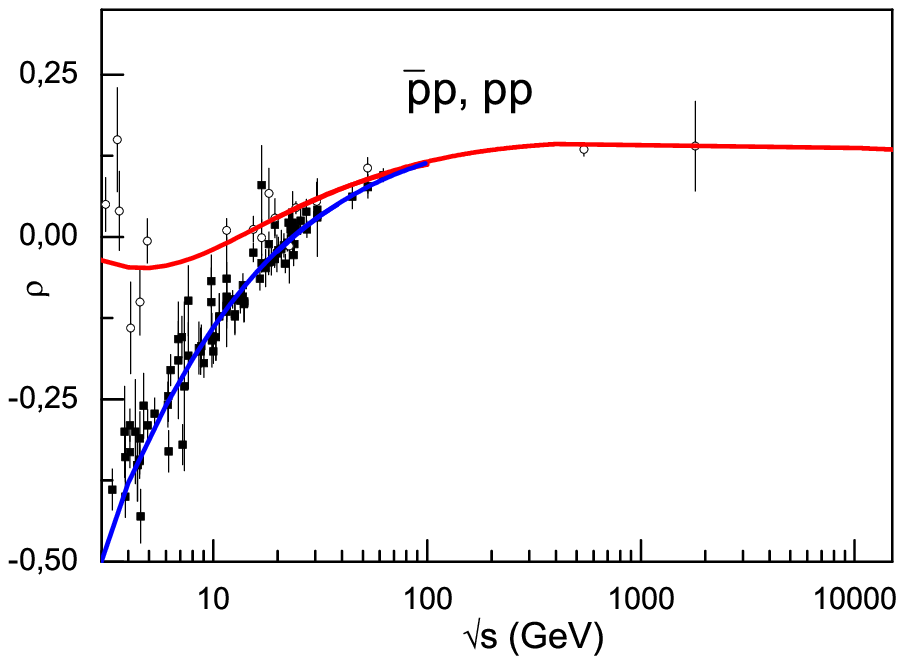}\\
        (a)\hspace{0.45\linewidth}(b)
}
\label{fig:Pomeron2}
%\vspace{0.5cm}
\caption{
(a) $pp$ and $\overline{p}p$ total cross sections calculated
from the model and fitted to the data in range $\sqrt{s}$ = 5 GeV--7 TeV.
(b) Ratio of the real to imaginary part for $pp$ and $\overline{p}p$ scattering amplitude calculated from the same model.
Upper curve - presents the $\overline{p}p$ calculation, lower curve - present $pp$ one.}

\end{figure}

%==================Figs.:  Figs.: =============================
\begin{figure}[fig:dif]
\center{
        \includegraphics[width=0.45\linewidth]{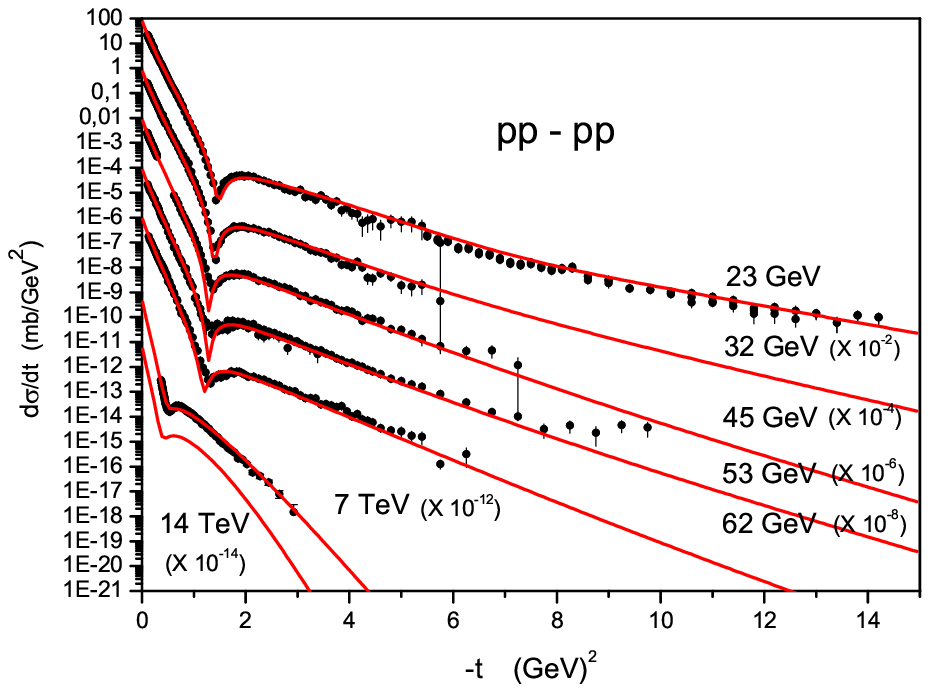}
        \includegraphics[width=0.45\linewidth]{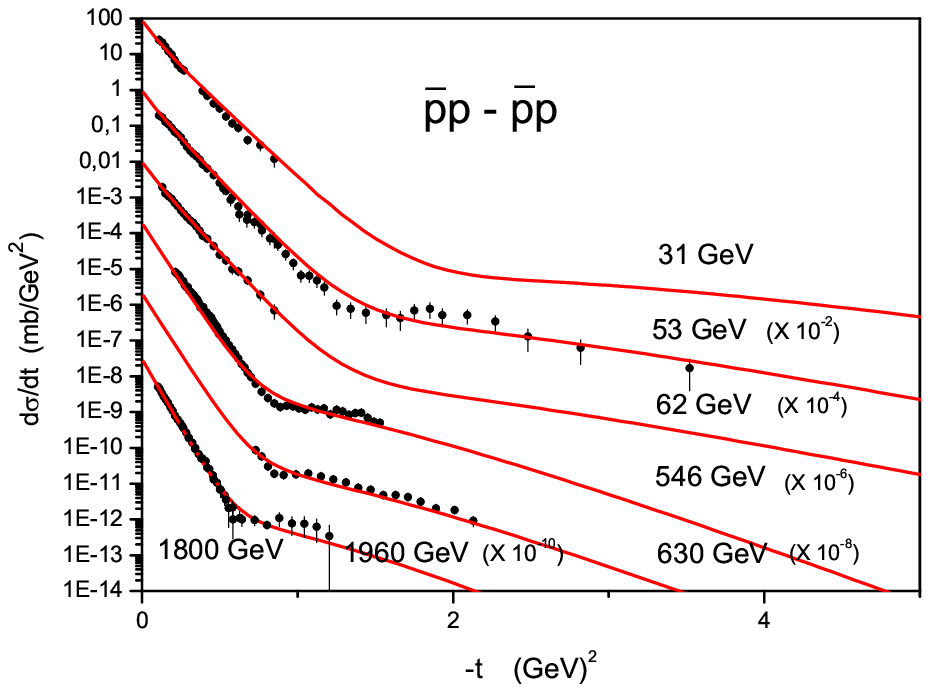}\\
        (a)\hspace{0.45\linewidth}(b)}
\caption{
%(a) Total $pp$ cross section calculated from the model, Eqs. (2-8,~\ref{eq:tr1}), without the Odderon term and fitted %to the data in the range $\sqrt{s}$ = 5 --- 30~TeV;
(a) Differential $pp$ (a) and $\overline{p}p$ (b) cross sections calculated from the model taking into account the spin and fitted to the data in the range $-t$ = 0.1 --- 15~GeV$^2$.
%(b) Differential $pp$ cross sections calculated from the model, Eqs. (2-8,~\ref{eq:tr1}), without the Odderon term and %fitted to the data in the range $-t$ = 0.1 --- 15~GeV$^2$.
%\label{fig:Pomeron}
}
\label{fig:dif}
\end{figure}

\end{document}